\newif\ifcomments\commentstrue
\newif\ifanon\anonfalse
\newif\iffullversion\fullversiontrue
\newif\ifsharingdraft\sharingdrafttrue

\iffullversion
    \documentclass[10pt,twocolumn,letterpaper]{article}
    \usepackage[margin=1in]{geometry}
    \usepackage{sectsty}
    \sectionfont{\fontsize{12}{12}\selectfont}
    \subsectionfont{\fontsize{11}{11}\selectfont}
    \renewcommand{\subsubsection}[1]{\subh{#1}} 
\else
    \documentclass[letterpaper]{article}

   \usepackage{times} 
   \usepackage{helvet} 
   \usepackage{courier} 
   \usepackage{graphicx}
   \usepackage{graphicx} 
   \usepackage{caption}
\fi

\usepackage[dvipsnames]{xcolor}

\usepackage{xspace}
\usepackage{pifont}
\usepackage{soul}

\iffullversion
    \usepackage[maxcitenames=3, maxbibnames=30]{biblatex}
\else
    \usepackage[maxcitenames=2, maxbibnames=2, url=false, doi=false]{biblatex}
    \AtEveryBibitem{
        \clearfield{journal} 
        \clearfield{booktitle}
        \clearfield{pages}
        \clearfield{address}
        \clearfield{month}
        \clearfield{isbn}
        \clearfield{editor}
    }
\fi

\usepackage{hyperref}
\usepackage[hyphenbreaks]{breakurl}

\ifcomments
    \newcommand{\sunoo}[1]{{\textcolor{violet}{S: #1}}}
    \newcommand{\peter}[1]{{\textcolor{blue}{P: #1}}}
    \newcommand{\ollie}[1]{{\textcolor{olive}{O: #1}}}
    
    \newcommand{\hide}[1]{\textcolor{orange}{#1}\xspace}
    \newcommand{\todelete}[1]{\textcolor{red}{#1}}
    
\else
    \newcommand{\sunoo}[1]{\ignorespaces}
    \newcommand{\ollie}[1]{\ignorespaces}
    \newcommand{\peter}[1]{\ignorespaces}
    
    \newcommand{\hide}[1]{\ignorespaces}
    \newcommand{\todelete}[1]{\ignorespaces}
    
\fi

\newcommand{\subh}[1]{\smallskip \noindent \textbf{{#1}}.}
\renewcommand{\paragraph}[1]{\subh{#1}}

\title{The Pitfalls of ``Security by Obscurity'' \\ And What They Mean for Transparent AI}
\author {
    Peter Hall \\ New York University\and
    Olivia Mundahl \\ New York University\and
    Sunoo Park \\ New York University\and
}

\date{}

\begin{document}

\maketitle

\begin{abstract}

Calls for transparency in AI systems are growing in number and urgency from diverse stakeholders ranging from regulators to researchers to users (with a comparative absence of companies developing AI). Notions of transparency for AI abound, each addressing distinct interests and concerns. 

In computer security, transparency is likewise regarded as a key concept. The security community has for decades pushed back against so-called \emph{security by obscurity}---the idea that hiding how a system works protects it from attack---against significant pressure from industry and other stakeholders\iffullversion~\cite{Bellovin2002Obscurity, Mercuri2003obscurity, Shipman2019opencode}\fi. Over the decades, in a community process that is imperfect and ongoing, security researchers and practitioners have gradually built up some norms and practices around how to balance transparency interests with possible negative side effects. 
This paper asks: \emph{What insights can the AI community take from the security community's experience with transparency?} 

We identify three key themes in the security community's perspective on the \emph{benefits of transparency} and their approach to balancing transparency against countervailing interests.
For each, we investigate parallels and insights relevant to 
transparency in AI.
We then provide a case study discussion on how transparency has shaped the research subfield of anonymization. Finally, shifting our focus from similarities to differences, we highlight key transparency issues where modern AI systems present challenges different from other kinds of security-critical systems, 
raising interesting open questions for the security and AI communities alike. 
\end{abstract}

\section{Introduction}

Artificial intelligence has proven to be highly impactful, with impacts in critical domains such as biosciences~\cite{Jumper2021AlphaFold}, health~\cite{Rajpurkar2022Medicine}, and public safety~\cite{Fine2024BailTrustAI}. It is also continually reaching new peaks in consumer and commercial interest, with foundation models such as Claude, Llama, Stable Diffusion, and the GPT family 
posited to be adaptable to diverse uses~\cite{meta2024LLM}. 
Despite their already widespread impacts, there is still much to understand about how recent developments in AI and machine learning models operate, and the broader human consequences their use and misuse may have. In the commercial sphere, companies have generally (though not always) 
preferred to hide their methods of data collection, training, fine-tuning, and other specifications~\cite{carlini2023extracting}.

At the confluence of these circumstances, 
many are invested in achieving (various notions of) \emph{AI transparency}. Researchers want to investigate what makes AI models, including production-line models, better or worse at various tasks (e.g.,~\cite{bender2021_stochastic_parrots}), and to understand unexpected or possibly harmful effects of model use (e.g., \cite{Awasthi2023Adversarial,Carlini2019Adversarial,nist2023riskframework}). Consumers, businesses, and other clients
should have assurances that companies selling AI-based tools are able to achieve what they are promising---and even before that, they should be able to understand what they are being promised. As increasingly consequential decisions are made with contributions from AI models, people impacted by these decisions need assurances about how AI impacts their lives (e.g., in healthcare \cite{Khan2023HealthcareAI}
\iffullversion
    , child custody \cite{Brooks2022AIFamilyMatters},
\fi
or bail setting \cite{Fine2024BailTrustAI}). Copyright holders want notice and fair compensation and attribution when others are using their works--often making large profits~\cite{NYTOpenAI2023, Proof2024YouTube}. Regulators want to understand and protect their constituencies from potential harms~\cite{weidinger2021ethical} while promoting innovation and competition in AI~\cite{FTC2024OpenWeights, FTC2024PNP, EU2024Speech}. Meanwhile, skeptics have variously argued that too much transparency will harm innovation, business interests (including trade secrets), data privacy, system security, national security/defense, and/or public safety (e.g.,~\cite{Patel2024security, Hosanagar2024regulating}).

An important emerging literature in both research and policy has responded to this ongoing debate by investigating transparency in AI\iffullversion---discussing possible definitions, implementation approaches, auditing frameworks, regulatory approaches, voluntary initiatives, and possible pros, cons, and impacts of all of the above\fi. 
Some have been trying to build in properties like explainability or interpretability to LLMs~\cite{Graziani2022interpretableAI}. Others are hoping to balance data privacy with open source models~\cite{FTC2024OpenWeights,Zuckerberg2024AI}. What this may mean is the subject of ongoing debate~\cite{Gibney2024NonOpen}. Others take a legal or policy approach, like the Biden Administration's Executive Order on Trustworthy AI and the European Union's AI Act~\cite{ExecOrder2023AI,EU2024AIAct}.

The idea that \emph{hiding how a system works 
protects it in some way}
is a key recurring theme in the current discourse on transparency in AI. This same idea is, however, one whose promises and pitfalls the security community is deeply familiar with~\cite{KahnCodebreakers}. Attempts to protect systems by hiding how they work have prompted the security community's participation in heated debates across academia, industry, and policy over decades~(e.g., \cite{Bellovin2002Obscurity,diffie2003risky,Mercuri2003obscurity,schneier2004nonsecurity, Shipman2019opencode}). We seek to understand the \iffullversion parallel \fi themes across the \iffullversion transparency \fi debates unfolding in AI and those longer entrenched in security.

Our paper thus offers a novel and systematic exposition of the essential principles underpinning transparency in security, which we believe have useful parallels to transparency in AI. 
That is, the central question of this paper is:

\begin{center}
    \emph{What insights can the AI community take from the security community's experience with transparency?}
\end{center}

\subsection{Transparency in Security}

In modern computer security, a hard-fought broad-based consensus has been established: {\it
Despite the intuitive idea that hiding a system should protect it,
transparency is often more beneficial for protection}. 
The consensus on this general principle is broad, though perspectives on how to implement the principle in specific contexts can be more varied.

It was not always this way. Before modern cryptography and systems security, even critical military communications often relied on \emph{security by obscurity} (e.g.,~\cite{Tohe2022codetalkers, KatzLindell}): namely, the idea, now well established to be fallacious, that hiding how a system works is an effective and adequate defense against adversaries\iffullversion ~\cite[1.3]{Ingeno2018Handbook},~\cite[2.1.1]{CryptographyEngineering}. \fi 

Moving away from security by obscurity has been a long and ongoing process.\footnote{Obscurity is still common in some contexts, e.g., in Digital Rights Management (DRM)~\cite{Liu2003DRM}.} 
Now, ``security by obscurity'' is a catchphrase in the security community with negative connotations, and relying upon security by obscurity is widely considered inadequate~\cite{Bellovin2002Obscurity}. 
Perfect unanimous agreement on the precise type of transparency that is best for every type of system and application context may never be reached---yet the received wisdom that transparency promotes better security practices and quicker mitigations of harm has held strong,
and served as a source of unity in the 
\iffullversion
    community~\cite{BurenvUS}.
\fi

We call this modern approach \emph{security by transparency}---a natural phrase at times used colloquially to describe the opposite of security by obscurity. As the security-by-transparency approach has developed through community experience and norm-building over time, a substantial part of it lies in folklore and institutional memory, written records of which are relatively sparse and scattered.

\subsection{Security by Transparency}

The notion of security by transparency predates computer security. As far back as the 19th century, Auguste Kerckhoffs~\cite{Kerchoffs1883crypto}, writing about how to design secure military communication, remarked that a truly secure system should remain secure even assuming the enemy knows everything about \iffullversion the design and parameters of \fi a system. That is, systems should \emph{by design} be robust even when their design is 
\iffullversion
    known.\footnote{This statement was originally applied to only a cryptographic/communication theory setting, though the concept now extends far beyond to security as a whole.} 
\fi

The argument for security by transparency is subtle: The idea that hiding how a system works can help protect it is not only intuitive, it is in a technical sense \emph{correct}. If you build all the best state-of-the-art security measures into your system and then choose not to disclose how it works at the end, it may in fact be harder for attackers to find vulnerabilities in the system, as they'll first have to figure out how it works. (And if you \emph{could} 
\iffullversion somehow \fi 
build a perfectly secure system, whether or not you disclose how it works doesn't matter; the security will be perfect either way. But 
\iffullversion
of course, 
\fi
we don't know how to build perfect systems; and the security community has embraced this reality as a critical engineering consideration \cite{Viega2001building}.)

Why, then, is security by obscurity so poorly regarded? 

First, \emph{relying} on obscurity for security is risky at best, and creates a false sense of security at worst. The phrase ``security by obscurity'' sometimes refers specifically to such \emph{reliance}: Having obscurity as the main protection, rather than as one measure among \iffullversion 
    many (as in the examples above).
\fi
Kerckhoffs' Principle warns directly against this. Second, decades of experience have shown that opacity leads to externalities that tend to indirectly undermine the robustness of designs and systems. 

We stress that security by transparency is a paradigm, not a rigidly defined set of rules. In a given situation, security by transparency provides principles with which to reason about what to disclose to whom, what to keep secret, and the possible consequences. These principles do \emph{not} prescribe disclosure of every last detail of a system, or prescribe any singular foolproof way to achieve security. 

\iffullversion
\subsection{Three Key Parallels}\label{ssec:five-key-things}

We highlight three parallel themes that we see as common to both the security and AI communities. 
We discuss these themes in greater depth in Sections~\ref{ssec:mainarg:modeling},~\ref{ssec:mainarg:vulnerabilities}, and~\ref{ssec:mainarg:interpretability}, respectively. 

\begin{enumerate}
    \item \textbf{Modeling is Essential For a Robust
    Understanding of Systems That Are
    Too Complex To Fully Model:} The security requirements of a given system may be too complex or arduous to fully describe, even for simple applications. As such, abstraction is required to help designers understand what a system can and cannot guarantee. Negative guarantees are as important as positive ones---\emph{threat models} help define what systems do \emph{not} protect against as well as what they do. Most modern AI tasks are likewise too complex to analyze completely. 
    This complexity likewise necessitates robust modelling and abstraction for both positive and negative guarantees.

    \item \textbf{``Many Eyes'' and Community Input:} 
    A key impact of transparent practices within security has been the ability to more efficiently and reliably address problems and mitigate harms. Three key transparent practices on which the security community has come to rely widely are: (1) inviting scrutiny to find weaknesses in systems; (2) accepting and incorporating feedback from such scrutiny (solicited or unsolicited); and (3) recognizing that wide community dissemination of such findings (e.g., through academic publications), subject to certain limitations to mitigate resulting harms, is a valuable contribution to security. These practices have long inspired controversy, and can still provoke tensions today. Some of the controversies around and arguments against these transparent practices have parallels in ongoing discourse about AI transparency.

    \item \textbf{Ubiquitous Technologies Necessitate
    Public Trust:} 
    Despite the fact that the average person is not expected to understand how encryption or AI work, transparent practices in security and AI benefit everyone if they are institutionalized and built in to infrastructure.
    This is especially critical for ubiquitous technologies that impacted individuals not only do not understand, but cannot meaningfully opt out of, such as banking and algorithmic decision-making.
    The security community's experience suggests that such circumstances may necessitate and support the widespread adoption of public trust mechanisms, \emph{even when} user understanding and market demand for transparency appears lacking.
    
\end{enumerate}

\fi

\subsection{Our Scope: \\ Transparency, Not Security}\label{ssec:scope}

Our focus is not on \emph{security} for AI systems but rather on \emph{transparency} for AI systems. We are interested \iffullversion primarily \fi in the parallels and potential for cross-pollination between the security and AI communities in their \iffullversion respective \fi debates around transparency. 

When we talk about security by transparency, our discussion will often center around how principles of transparency promote \emph{security} in some way---naturally, as the context is the field of security. 
We argue, however, that these discussions' relevance to the AI community extends beyond securing AI, to AI transparency much more broadly---for at least two reasons.
First, we are focused on high-level principles, and security at a high level is simply about ``building systems to remain
dependable in the face of malice, error, or mischance'' ~\cite{Anderson2008Security}---that is, the very same concerns underlying many of the calls for transparency in AI mentioned above, including some we might not usually think of as ``security concerns.''
Second, transparency aids security \emph{as a discipline} in ways that may translate to other disciplines: For example, allowing researchers to uncover more efficient solutions and richer properties and functionalities of systems, and to better understand system guarantees and limitations. 

\subsection{Structure of Paper}

We provide an overview of transparency in security and a survey of transparency in AI (Section~\ref{sec:background}), and then present our three key parallels in depth; for each parallel, we present (1) a specific perspective on transparency in security, to (2) what aspects may be similar in AI, to (3) parallels with AI that may inform thinking about potential benefits and approaches to transparency in AI research and practice (Section~\ref{sec:mainarg}). 
Then we explain how our discussion of transparency in AI complements the existing literature and AI transparency goals (Section~\ref{sec:how-transparency-complements}), and offer a case study of how norms around transparency have developed and evolved over time in one sub-field of security: Anonymization (Section~\ref{ssec:casestudy:deanon}).
Finally, 
we highlight areas where AI presents novel challenges we believe not present in the security context, including directions for future research (Section~\ref{sec:differences}), and
we discuss common arguments against transparency which we believe have parallels in security (Section~\ref{sec:discussion}).

\section{Background}\label{sec:background}

\iffullversion
\subsection{Transparency in Security}
\fi\label{ssec:background:security}

    By the late 19th century, cryptographers had reservations about security by obscurity. In the seminal 1883 work ``La Cryptographie Militaire''~\cite{Kerchoffs1883crypto}, Kerckhoffs voiced six principles he believed essential to strong security systems. Among them, known now as Kerckhoffs' Principle, stated approximately: ``The system must not require secrecy and can be stolen by the enemy without causing trouble.''\footnote{In the original French: ``Il faut qu'il n'exige pas le secret, et qu'il puisse sans inconvénient tomber entre les mains de l'ennemi.''} Thus began the counter viewpoint, security by transparency. 
    
\iffullversion
Kerckhoffs' idea was later reframed by Shannon in what is now known as Shannon’s Maxim: ``The enemy knows the system being used''~\cite{Shannon1949secrecysystems}. This made explicit the link between transparency and provable security, which underpins all cryptographic research now. 
\fi

\iffullversion\paragraph{Disclosure Practices} \fi From the earliest days of consumer computer systems, researchers and concerned citizens have been finding and sharing found vulnerabilities. In the late 1980s and '90s, this was through email chains and online zines like Bugtraq\footnote{\url{https://seclists.org/bugtraq}.} and Security Digest,\footnote{\url{securitydigest.org}.}; now, security experts are more likely to communicate vulnerabilities to the developers through formal bug bounty processes and to the public through peer-reviewed publications. 

There has been much debate over the best way to disclose vulnerabilities both to developers and the public. The details of this debate are beyond our scope;\footnote{There is plenty of literature on the topic (e.g.,~\cite{Moura2023Disclosure, Wicker2021zeroday, CERTGuide}}
here, we just summarize a few key terms related to vulnerability disclosure. 

\emph{Full disclosure} involves sharing all information about an attack with the public immediately. 
\iffullversion
    Many are concerned that full disclosure could result in a situation where attacks would be exploited widely~\cite{Culp2001InfoAnarchy}, as it can result in public knowledge of a possible attack before mitigations can be put in place.
    \emph{Coordinated vulnerability disclosure} (CVD) involves an initial private disclosure followed by publication after a delay to allow for fixes---the initial disclosure is regarded as a best practice to enable mitigations before a vulnerability is widely known, and the eventual public disclosure is widely regarded as a best practice to inform all affected stakeholders~\cite{CERTGuide}.\footnote{The exact details of appropriate CVD timeline are context-dependent and a subject of ongoing community discussion.}
    
\fi
Increasingly, many organizations have groups who handle bug and vulnerability reporting,\footnote{See, e.g., the Microsoft Security Vulnerability Research group.}; this streamlines CVD processes and timely fixes. \emph{Bug bounty programs},  where community members are invited to submit found vulnerabilities to the company 
\iffullversion
    in exchange 
\fi
for money or perks, are a way for organizations to proactively seek disclosures.\footnote{See, e.g., Bugcrowd: \url{https://www.bugcrowd.com}.}

    \iffullversion\paragraph{Attacks and Vulnerability Papers} \fi At security venues, attack papers 
    \iffullversion
    (also called vulnerability papers)
    are a common sight now, though they were more controversial in the past before their research value was 
    as well understood
    as it is today.
    
    \fi
    Basin and Capkun~\cite{Basin2012attacks} provide one account of the case for attack papers: Finding and fixing bugs, learning flaws which may guide research, understanding compatibility or lack thereof, finding exact security guarantees. \iffullversion
    Basin and Capkun highlight that good attack papers provide insight beyond the attack (say, through suggested or implemented patches).
    We refer to their discussion for more detail.
    \fi

    \iffullversion
    
    We make two observations as to the relationship between these works and transparency. First, the existence of the flaw in a system is the vulnerability, \emph{not} the specific attack found. That is, choosing to hide a flawed system does not wholesale protect it from some malicious party being able to find and exploit said flaw. Especially in the case of sensitive or valuable data, understanding and providing mitigations for these vulnerabilities actually helps protect more than solely relying on hiding the flaw. Second, publishing a vulnerability (again, with additional insight), promotes better design. That is, understanding and in particular disseminating different sources of vulnerabilities not only pushes the designers of a flawed system to fix their own, but it also allows others to learn what \emph{not} to do in the future. 
    
    In these ways, transparent discussion of attack papers promotes better practices. However, simply describing an attack is not sufficient. Not only is providing additional insight and mitigations a generally accepted key part of writing an attack paper, but the security community also generally sees the value in a graded disclosure process, rather than immediate full (public) disclosure.

    \fi

\iffullversion
\subsection{Transparency in AI}\label{ssec:AITransparencyBackground}
\fi

\subsubsection{Calls \& Goals for Transparent AI}
Calls for transparency in AI are numerous
(e.g., \cite{White2024MOF}). 
Notable relevant terms and areas of study include:
explainability~\cite{NIST2023Framework}, interpretability~\cite{Rudin2019InterpretableAI,NIST2023Framework}, accountability~\cite{Novelli2023Accountability, NIST2023Framework}, trustworthiness~\cite{NIST2024RespAI, NIST2023Framework}, and robustness~\cite{EU2021HarmonisedAI, NIST2023Framework}.\footnote{The nuances and precision of the definitions of these terms vary in the literature, and are beyond our scope. (See, e.g.,~\cite{Lipton2016Mythos} for a critique of interpretability notions.)} Recent advances are making important progress in AI modeling~\cite{Dev2024BuildingGuardrails} but often focus on defining largely positive guarantees. 
Additionally, some risks can be difficult or elusive to define within the scope of rigorous models~\cite{Polemi2024CurrentRiskAI}. 
Disclosure practices are being developed~\cite{DoJAIDisclosureGuidelines, Cattell2024flaw}, but AI researchers remain at risk due to uncertainties and discrepancies regarding policies~\cite{Tiku2024HinderResearch}. Prominent motivations for these efforts towards transparency include public trust~\cite{Andrada2022agency,schneier2023publicai}, technological advancement~\cite{fukawa2021dynamiccapability,Bostrom2018AIStrategy}, mitigation of bias~\cite{Carlson2017Transparency, Ferrara2024BiasMitigation}, and legal liability~\cite{Novelli2023Accountability, Cooper2022Accountability}.

\subsubsection{Types of Access} Many AI or ML systems using deep learning algorithms cannot be deciphered or broken apart~\cite{Eschenbach2021BlackBoxProblem} and instead are reasoned about solely through input-output behavior (known as \emph{black-box} access).  
Proponents of black-box models argue that they perform more accurately and effectively~\cite{Loyola2019BvWBox}. Critics warn that mitigations or design changes may be harder to find or analyze~\cite{Bathaee2018BBoxCausation, Rudin2019InterpretableAI}.

Post-hoc and ante-hoc solutions have been proposed to better understand AI models~\cite{Zednik2019BBoxProblem}. 
Post-hoc solutions are often linked to Explainable AI or XAI, developed to increase the capacity for human understanding of AI models ~\cite{Hassija2023XAI}. 
Conversely, ante-hoc solutions investigate how to make an AI system transparent from the beginning ~\cite{Felzmann2020TbyDesign}. Ideas range from sharing the model itself, the training data, the weights of the model, and more ~\cite{Andrada2022agency,Eschenbach2021BlackBoxProblem,White2024MOF}, but there is little consensus as to which types of access are necessary or sufficient.

\iffullversion
    \subsection{Open Source, Security, and AI}\label{sec:opensourcesecurity}
    
    Discourse around open source software encompasses politics, philosophy, the law, and many other domains. 
    
    Most software uses open source code somewhere in its stack, and the Cybersecurity and Infrastructure Security Agency notes that open source software even fosters innovation by allowing developers to pool their knowledge~\cite{CISA2023roadmap}.  
    Many of the most-contributed open source projects are commercially backed, despite the fact that this work is open to anyone including competitors. However, it is important to note some caveats to open source. Schneier~\cite{Schneier1999opensource} has noted that open source can only increase security if people are actually studying the code, and if developers are prompt in bug fixing. However, Sharma~\cite{Sharma2022digitalcommons} claims that the decentralized nature of open source may introduce adverse incentives toward this, as those who benefit from open source software are not proportionally incentivized to develop it further.
    
    So-called ``open source AI'' has seen great interest in the study of transparent AI, with companies like  Meta~\cite{Touvron2023Llama2}, Huggingface~\cite{Almazrouei2023Falcon}, AI2~\cite{Walsh2024OLMo}, and EleutherAI~\cite{Biderman2023Pythia} developing their own open source models.
    At the same time, there are concerns that the term lacks a well-defined meaning 
    ~\cite{Gent2024OSAIMeaning}. Experts greatly disagree on the definition of open source AI, and some have warned of ``openwashing''~\cite{Liesenfeld2024OpenWashing}, where developers claim open source without following accepted principles of the community.  
    It may be beneficial for larger AI companies to be \emph{perceived} as open source~\cite{Bostrom2018AIStrategy}. 
    The standards body Open Source Initiative~\cite{OSI2024Def} is currently working on developing a concrete framework for open source AI with input from community members~\cite{OSAI2024Def}.
\fi

\section{Connections In Transparency: \\ From Security To AI}
\label{sec:mainarg}

\iffullversion
    This section highlights three key themes underpinning the security community's approach to transparency.
    For each, we discuss parallels with AI that may inform thinking about the gains transparency would bring to AI research and practice. 
\fi

\subsection{Modeling is Essential For a Robust Understanding of Systems That Are Too Complex To Fully Model}
\label{ssec:mainarg:modeling}

Perhaps paradoxically, modeling serves as an essential tool to maintain a robust understanding of what we can \emph{and cannot} guarantee about the performance and failure modes of systems that are too complex to fully model --- which have (security) requirements too complex to fully model. 

But wouldn't modeling such a system (and its requirements) be futile as any model would fail to capture important aspects of real-world operation? Even worse, wouldn't modeling such systems rather lead to dangerous oversimplifications and misunderstandings? These are reasonable questions: oversimplified models do suffer from these serious drawbacks \emph{when their limitations are not understood}. 
In contrast, the type of modeling for which we advocate critically \emph{includes} clear characterization of the model's own limitations: that is, modeling what we can \emph{and cannot} guarantee about a system on the basis of a given model. 

Security engineering involves building systems with stated \emph{threat models} and \emph{assumptions}, which delineate (1) what functionality, confidentiality, and robustness guarantees a system is designed to have (2) against which kinds of threats (3) under what conditions and, implicitly or explicitly, (4) which kinds of functionality are \emph{not} guaranteed and what kinds of threats are \emph{not} protected against. 
Threat modeling has proven an essential technique to build (necessarily) imperfect systems that have clear specifications of \emph{what is guaranteed} and \emph{what is not}, and in what scenarios these guarantees are assured.
The threat model is generally considered part of the system specification: it is seen as necessary to understand ``what's in the box,'' as well as how (not) to use it and for which application contexts it is appropriate.

Transparency plays a key role here in several ways. First, if an assumption necessary for a system's security is proven incorrect, clearly stated assumptions allow for the community to quickly adapt by pruning ideas which rely on 
\iffullversion      
    it.\footnote{For a recent example in cryptography, see the break of SIDH~\cite{castryck2023sidh}, in which the authors note in the abstract that their attack extended to a leading post-quantum candidate protocol SIKE.} 
\fi
Second, by making these assumptions explicit, we encourage research to find ways to test or bolster the assumption,\footnote{E.g., the fallibility of passwords in practice gives rise to other authentication schemes like biometrics, 2FA, etc.} as well as make it more likely that the research community will 
\iffullversion
    find out in a timely fashion
\fi
if an assumption does not hold. Third, it enables both developers and users to plan ahead for system failures in a way tailored to the threat model. Fourth, it aids stakeholders to evaluate in what contexts the use of a tool is riskier, and in what contexts not to use a tool at all. Finally, 
\iffullversion
    clearly stated assumptions and threat models,
    while imperfect models of reality, 
\fi
serve as stepping stones for further research and more refined models that further enhance the above benefits. 

All of the above points have baked into them the implicit idea that we cannot ``simply'' build a system to be secure, not only because we do not know how to build perfect systems but because we cannot write a single specification of what ``security'' 
\iffullversion
    is and how to achieve it. 
\fi
The objective of security is a highly context-dependent, and fully specifying what ``security'' should mean is often intractable.
In sum: (1) there is no general-purpose certification of ``secure enough'' with which we could label any system that meets a fixed set of requirements; (2) we aim to design specific threat models appropriate to limited contexts, stress they are \emph{not} general purpose, and clearly define what they do \emph{not} cover; and (3) we associate clearly stating and disseminating such threat models with the benefits above.
\iffullversion
    We believe this 3-step story has some parallels in the AI context.
\fi

\paragraph{Connection to AI}
Some simple AI 
\iffullversion
    tasks, just like some simple security tasks,
\fi 
are relatively clear and self-contained in their goals. 
But in many \iffullversion of the more interesting \fi settings\iffullversion{ in both AI and security} \fi, 
the goals can often be less clear, expressed as heuristics, inherently incomplete, inherently sociotechnical (and thus not amenable to complete technical specification), and/or context-dependent.
Explicit ``threat'' modeling in the AI context can embrace this incompleteness and the idea that \emph{these systems and requirements are too complex to fully model}, and seek to delineate what is not clear, what is incomplete, and what can(not) be guaranteed. By doing so, it has the potential to play a similarly critical role in reinforcing robustness for AI systems deployed in critical contexts when robustness is a context-dependent moving target.

None of this is to say that AI research is devoid of modeling---far from it. Recent positive strides include theoretical modeling (e.g., \cite{Górriz2023ComputationalApproaches}), candidate definitions of robustness, fairness, and privacy (e.g., \cite{NIST2023Framework}), transparency in experimental design (e.g., \cite{Felzmann2020TbyDesign}), and transparency of societal impacts (e.g., \cite{NIST2024RespAI}). These efforts make important progress on defining \emph{positive guarantees} but we have seen less depth of engagement with \emph{negative guarantees}: Precisely characterizing the types of situations and failure modes \emph{not} addressed by proposed models and definitions. Here, we mean going beyond the important initial step of mentioning that certain enumerated things out of scope of one's model; we mean, rather, striving towards a \emph{complete} characterization of what's not in scope --- such that if anyone thinks up a threat, they can straightforwardly ascertain whether and how it is in scope, by reading a concise 
\iffullversion
    ``threat model'' section of the system specification.
\fi
We recognize that this is a complex goal that will not be achieved overnight, and that important research has already been done in this direction (e.g.,~\cite{Stadler2024limits}); our aim is not to criticize but to highlight this natural and fruitful avenue for future progress.

\subsection{``Many Eyes'' \& Disclosure}
\label{ssec:mainarg:vulnerabilities}\label{ssec:mainarg:disclosure}

A key impact of transparent practices within security has been the ability to more efficiently and reliably correct errors, vulnerabilities, and other problems. Examples of such transparent practices include bug bounty programs, cryptographic standardization processes, and the publication and recognition of attack papers. These practices all invite scrutiny from all over, letting the entire community contribute. Underpinning this reasoning is what is known as the ``Many Eyes'' Theory, also known as Linus's Law: ``Given enough eyeballs, all bugs are shallow''~\cite{Cathedral1999}. That is, giving all stakeholders the ability and incentive to investigate a system aids in better design and mitigation. 

\iffullversion Finding and patching bugs and vulnerabilities to prevent attacks is a core part of security research (as also noted in \iffullversion Section~\ref{sec:background}). \else the Background). \fi In academia, this may take the form of vulnerability papers discussing and finding mitigations for attacks in systems. In industry, this may instead consist of internal security reviews and red-teaming exercises, as well as hiring expert third-party auditors\footnote{E.g.,Trail of Bits.} --- and often, committing in advance to release their report publicly.\fi

Positively receiving and proactively seeking community input on security issues are \iffullversion also \fi essential practices that have become much better established over time, although these were rare practices in earlier days (and are far from ubiquitous even today; progress is ongoing). \iffullversion In an early inspiring example, \fi RSA~\cite{RSA1991Announcement} famously prompted the community to find weaknesses in their systems and share that information for prizes (eventually claimed, like in~\cite{Cavallar2000Factor}). Now, many companies solicit vulnerability reports or offer ``bug bounties'' for 
\iffullversion
    them.\footnote{Note that not all bug bounty programs may be in good faith or truly protective of researchers~\cite{Albergotti2021bounty,ElazariBugBounties}, \cite[\S6]{SpecterKW20}. See \href{https://www.bugcrowd.com/bug-bounty-list/}{Bugcrowd} for a list of bug bounties.} 
\fi 
Organizations that set standards and offer community and support for soliciting reports are also influential and growing.\footnote{E.g., \href{https://disclose.io}{disclose.io}, \href{https://hackerone.com/bug-bounty-programs}{Hackerone}, and \href{https://www.bugcrowd.com}{Bugcrowd}.} These recent trends
\iffullversion
    starkly contrast with many
\fi
early examples of hostile reactions to reports of security issues 
\iffullversion
    (e.g., ~\cite{MBTAvAnderson,WeevCase,McDanelCase}), although hostile reactions still do happen (e.g.,~\cite{SpecterKW20}).
\fi

\iffullversion
Cryptographic standardization efforts by NIST (e.g., ~\cite{NIST2016Call}), which consist of multi-round competitions where candidate protocols are required to have fully public descriptions~\cite{NIST2017Requirements}, are another powerful example of proactive seeking of community input. In these efforts, the security community at large is invited to evaluate the security of all candidates, and vetting new cryptographic tools with this kind of public testing has become an essential best practice before deployment at scale---even when those cryptographic tools end up deployed in industry contexts where broader system designs are kept secret.\footnote{NIST in particular has also prompted similar community input in the context of AI~\cite{nist2023riskframework}.}. 

Efforts like the above have become widely accepted because of their potential to find and mitigate security flaws faster and more effectively by leveraging community input. \fi

Realizing the full potential benefit of community input is challenging in practice as it requires trust and mutual understanding between stakeholders with incentives that are not always aligned: the community providing input and the community receiving input (though these are not always disjoint). Lack of established precedents and collaborative norms can lead to uncertainty and tensions, as both the security and AI communities have experienced---for security, more acutely in earlier decades. These, in turn, impede constructive progress based community inputs. The security community has, over time, developed stronger collaborative norms around disclosure processes that facilitate mutual benefits from community inputs; that said, the process is imperfect and norm-building is very much ongoing.

When some flaw in a system is found, choices must be made about who to tell when (if at all). 
While specific standards vary, it is a widely held belief within the security community that there should generally be some form of a graded disclosure process rather than a single public 
\iffullversion
    disclosure,\footnote{It is far from a closed discussion on when or whether the risk of public disclosure is worthwhile, see for example the DEFCON Ethics Village workshops: \href{http://ethicsvillage.org/2018.html}{http://ethicsvillage.org/2018.html}.}
\fi but that public disclosure \iffullversion of some kind \fi should eventually happen and generally promotes security. 

When a report of a flaw is received, choices must likewise be made about how to respond and how to address the report (if at all). Only when the one side's approach to who to tell when and the other side's approach to how to respond and address the report are aligned can the full potential benefit of community input be realized. And yet, incentives are not always aligned between the two sides.

\iffullversion
    As further contextualized in the Background (discussing CVD), increasing 
\fi
community norms around disclosure processes have made  \iffullversion important partial \fi progress toward aligning the two sides' 
\iffullversion
    approaches (e.g.,~\cite{CERTGuide, gfcrc2022, AskAnEthicist}).
\fi
The community's acceptance of and collaboration on vulnerability reports today stands in notable contrast to earlier decades, where security researchers were more often subject to aggressive legal and reputational attacks in reaction to their findings~(e.g., \cite{Rauch2022Charlie, MBTAvAnderson, Garrido2017legalrisks}). The difference between research that is beneficial for security and \emph{maliciously attacking} a system---confusion over which was a recurrent feature earlier on---became better understood and more clearly distinguished over time, by stakeholders in both industry and research (e.g., \cite{Rauch2022Charlie}).
Avenues for coordinating disclosure continue to expand, e.g., involving an intermediary such as a government body~\cite[\S6]{SpecterKW20}.

That said, much progress remains to be made. Terms of service often still prohibit security researchers from accessing and analyzing systems for research 
\iffullversion
    purposes---sometimes, even the terms for participation in bug bounty programs include such restrictions~\cite{ElazariBugBounties}.
    Security researchers continue to face legal and reputational attacks over their research, and navigating disclosure processes can still involve tensions and hostility~\cite{Park2024legalrisks,Moura2023Disclosure,SpecterKW20}.
\fi 

\paragraph{Connection to AI} At present, there are many barriers to implementing similar processes in AI. The terms of service of many prominent AI tools disallow or disincentivize the work needed to study vulnerabilities and the disclosure of such flaws~\cite{Longpre2024RedTeaming}. AI researchers have been targeted for their research by developers; some have been instructed to redact certain parts of their research or findings~\cite{Carlini2024Stealing}. These barriers feel very familiar to the experiences and history of the security 
\iffullversion
    community, and in fact many have explicit parallels.
\fi
And in security, it is these early barriers that led 
\iffullversion
    eventually to the disclosure processes we witness today.
\fi

To ensure that AI systems are evaluated\iffullversion, benchmarked,\fi 
~and properly understood, the security community's experience suggests that promoting an active research community and access by the research community can in fact promote a more robust end product with better understood guarantees. \iffullversion On the other hand, neglecting the research community does not mean that it will go away.\fi
\iffullversion

In addition,
\fi
a clear, codified disclosure process could go a long way to protecting any private information while also increasing trust between the community and developers. Some have attempted to explore what this could look like ~\cite{Rando2022RedTeaming, Longpre2024RedTeaming}, and we believe this is critical to transparency for AI. Of course, as the security community's experience also illustrates, building community around such processes takes time.

\iffullversion
    \subsection{Ubiquitous Technologies Necessitate Public Trust Mechanisms}
\fi
\label{ssec:mainarg:interpretability}

A common concern about transparency (in security and AI) is that average users cannot generally be expected to understand the technologies involved even if they are carefully explained in full detail---so then, what purpose does the transparency serve? Laudable efforts have been made to promote understandability for lay 
\iffullversion
    users,\footnote{See, e.g.,~\cite{Zhang2021Multimedia} for a survey of efforts in security.}
\fi 
but the full complexity---and the range of risks and failure modes---of these technologies will not realistically be understood by every user impacted.

Despite the fact that the average person is not expected to understand security best practices or how encryption works, transparent practices in security benefit everyone. Whether they are aware of it or not, everyone who uses computers uses security features and engages in compliance with security best practices---they are built in at a hardware level, at a protocol level, and at a networking level. This is facilitated by transparent practices allowing for users and developers to buy in to best practices. The security community's experience shows that the ubiquity and broad impact of systems where security matters necessitates and supports the widespread adoption of public trust mechanisms, \emph{even when} user understanding and market demand for transparency appears lacking, and companies seem to lack incentives to be transparent.
The result is more reliable and coordinated security practices across complex interconnected systems.

Many commonplace interactions online require users to rely upon the safety and robustness of systems where security matters. By using these systems, users place implicit trust in the systems; and the more essential the use of the system is to modern life, the less choice they have about placing their implicit trust in the systems. And sometimes, the choice to use the system in a way that impacts a given person is made by someone else (e.g., an employer).

Thus, people implicitly trust that their credit card information will not be stolen when they enter it in an online form. They trust that their messages are private and go to the intended recipient. They trust that their webcams are not on when they're not using them, and they trust that their smart watch monitoring their health condition will flag anomalies.
They trust that their employer's payroll system works.
Such trust does not come directly from individuals studying transparent literature to understand design practices, but rather, indirectly through public trust mechanisms. 

By public trust mechanisms, we refer to a host of approaches. For example, in fields like civil engineering, certifications and regulation serve as public trust mechanisms. In biomedical sciences, extensive testing is another public trust mechanism. Fields like psychology and physical sciences have academic standards and checks as public trust 
\iffullversion
    mechanisms.\footnote{These are neither mutually exclusive nor exhaustive. See Section~\ref{sec:related} for further discussion.}
\fi
In human subjects research, Institutional Review Boards (IRBs) and the Belmont Report's codified ethical framework serve as public trust mechanisms~\cite{BelmontReport, Stark2011IRB}.\footnote{The Menlo Report~\cite{MenloReport} was developed, explicitly based on the Belmont Report, to similarly provide ethical principles for computer science research. While it has had important impacts, the Menlo Report does not (yet) appear to have had the scale of impact that the Belmont Report has respectively had~\cite{Finn2023Menlo, Stark2022HumanSubjects}.}
Some of these mechanisms are legal or regulatory; some are institutional, professional, or based on community norms.

In these cases, the stakes are clear: People could die if things are not done safely. 

When considering the safety and robustness of complex computing systems, though, the potential harms often seem more 
\iffullversion
    intangible.\footnote{Of course, people \emph{can} die, and have in fact died, from malfunctions in security-critical systems including AI systems (e.g.,~\cite{Long2009SoftwareGlitch, AP2022CarCrashes}). However, more commonly, harms from computer system failures tend to provoke more of a ``generalized feeling of unease'' rather than ``blood and death, or at least of broken bones and buckets of money,'' which tends to feel much less ``immediate and visceral'' even when downstream harms may in fact be widespread and life-changing~\cite{Bartow2006PrivacyLaw}.} 
    Even in a case where someone recognizes an issue or breach, it may not be as easy to understand the possible consequences and resulting harms, or to understand the actions that can be taken to prevent such failures.
\fi 
This lack of visibility can lead to a lack of awareness and proactivity among lay users as well as a lack of incentive for companies to improve either security or transparency---in economic terms, creating a \emph{moral hazard}
\cite{Vagle2017moralhazard}.

These circumstances may indicate that legal or regulatory intervention to promote public trust, and to promote transparency more broadly, could have significant benefits.\footnote{Examples from the security context include measures such as the U.S. Cybersecurity and Infrastructure Security Agency's (CISA's) facilitation of information sharing around vulnerabilities~\cite{CISAInfoSharing}, and NIST's facilitation of the publication and scrutiny of system designs (see Section~\ref{ssec:mainarg:disclosure}).}

We see strong transparent practices as a way that security researchers have built up community-based public trust mechanisms, that in turn promote systemic deployed security that benefits everyone notwithstanding all of the countervailing forces and considerations discussed above.

\paragraph{Connection to AI}
The ongoing debate around AI transparency is likewise grappling with the facts that the average user cannot be expected to understand AI or fully evaluate the benefits and risks associated with its use, that companies generally seem to lack incentives to be transparent, and that users \iffullversion generally \fi do not seem to create market demand for transparency\iffullversion---yet increasing advocacy for transparency suggests that transparency would provide broad societal benefits and would require systemic deployment\fi.

AI is also growing rapidly, being deployed in a wide range of contexts with broad impacts, and is poised to be embedded into \iffullversion everyday \fi technologies at a scale that leaves users little choice on whether to engage with it. In many cases (including, e.g., policing~\cite{Angwin06Bias} and welfare distribution~\cite{Zouridis2020welfare}), individuals cannot opt out of algorithmic decision-making, including in cases where these may have serious consequences. 

Indeed, many have taken notice of the prevalence of AI in their lives, and many have raised concerns \iffullversion over copyright~\cite{He2024beasts}, plagiarism~\cite{NYTOpenAI2023}, rogue machines~\cite{Hendrycks2023Catastrophic}, fidelity~\cite{bender2021_stochastic_parrots}, labor rights~\cite{Acemoglu2022AIJobs}, and competition~\cite{Hall2024norms}\fi. However, the incentives are not necessarily aligned for any one group to contend with these issues, even if resolutions are in the public's interest. \iffullversion For example, companies developing AI and AI-powered tools have had mixed results in meeting the concerns listed above while largely maintaining a closed or black-box approach (e.g, ~\cite{Renieris2024Compliance, Heikkilä2024SelfRegulation}).\fi
\iffullversion

\fi
The combination of this kind of \emph{ubiquity and impact} with the \emph{lack of understandability and tangibility to users} \iffullversion(leading to a seeming lack of market demand)\fi, and related \emph{weak incentives for companies} creates the conditions---shared between AI and security---that we believe necessitates transparent practices within community-wide public trust mechanisms. These, in turn, promote systemic safety and robustness measures that can benefit society despite the complex incentives involved.

We believe transparency is necessary as a public trust mechanism for secure systems and robust AI alike, both for the benefit of the public and also for the adoption and continued use of these technologies to their full potential. 

\iffullversion
    \section{How Transparency Complements \\ Existing AI Research Goals} 
    \label{sec:how-transparency-complements}
    
    In the previous section, we described three themes for how transparency informs security and security research: (1) modeling as a tool for understanding guarantees of complex systems; (2) vulnerability reporting and disclosure norms as a tool to promote good practices and trust; and (3) transparent practices as public trust mechanisms in settings where public trust is needed. 
    
    We believe these themes are resonant to both the security and AI research communities. While the history of security research is evidence enough for the former, next, we discuss how these distilled ideas are complementary to existing AI research directions.
    \begin{enumerate}
        \item \emph{Modeling:} Our focus on discussing modeling in security is to point out the role of threat modeling in security in explicitly mapping out both what a system can guarantee and what it cannot. This latter goal is a critical aspect of security modeling, both because it narrows the scope of attacks to be concerned about, and because it defines areas for improvement and new techniques. Recent works (see Section~\ref{sec:background}) are making important progress in formally modeling AI, though with more focus on defining largely positive guarantees; we have seen less depth of engagement with negative guarantees. 
        \item \emph{``Many Eyes'' and Disclosure:} The AI community is starting to develop disclosure processes, with difficulties that we see as mirroring the history of disclosure in security. The AI community may consider drawing upon security's prolonged experience for parallel institutional knowledge on both mistakes and successes. We see protecting researchers as of critical importance here---that is, we believe norms going forward should center on protecting those that are ethically disclosing sensitive information about unexpected system behavior that could lead to large-scale harms, whether through deliberate exploitation or otherwise. Researchers continue to face risk in both fields due to remaining uncertainties around disclosure (e.g., \cite{Park2024legalrisks, Longpre2024RedTeaming}); 
        this is a key reason to establish good disclosure processes as quickly as possible.
        \item \emph{Public Trust:} The ongoing history of security research has made clear that transparency in systems is not only a useful property but in fact an indispensable best practice, although it was not always viewed as such. There are many researchers in AI who agree with this view and have been advocating for a broader consensus on the same in the AI context (e.g.~\cite{Linardatos2020XAI, Hassija2023XAI,}).
    \end{enumerate}
    
\fi

\section{A Case Study on \\ (De)-Anonymization}\label{ssec:casestudy:deanon}

\iffullversion 
We have discussed several ways in which transparency has gradually been normalized in the study of security. We emphasize again that these lessons were not self-evident at the dawn of the field, but rather they developed over time implicitly. Here, we look at how the themes highlighted in Section~\ref{sec:mainarg} have featured in transparent practices within one subfield of security, in this case \emph{anonymization}. The goal of this case study is to give a more concrete view of how transparency has developed, as well as how it has resulted in gains for the community.
\fi

Anonymization refers to methods designed to hide individually identifiable information in a dataset while retaining useful data accesses and statistics. 
For some time, anonymization was pitched as the answer to being able to make use of the vast datasets of private information that online platforms and governments amass (e.g., communications, content consumption, or census data), while also protecting individual safety and 
\iffullversion
    privacy~\cite{Ohm2009Anonymization, Rubinstein2016AnonRisk}---an appealing prospect of having your cake and eating it too. 
\fi

Proposed anonymization techniques were taken by some as a green light to ``anonymize'' sensitive datasets and then use them for any purpose, including publishing them. 
\iffullversion
    For example, ``anonymized'' census data~\cite{Sweeney2000Simple}, Netflix histories \cite{Narayanan2008deanon}, and student educational data~\cite{Cohen22} were posted online.
\fi
A series of works around the 2000s exposed this as overly optimistic~(e.g., \cite{NYT2006AOL, Narayanan2008deanon, Sweeney2000Simple}), de-anonymizing willingly published user data from sources such as AOL, Netflix, and population registers\footnote{See https://latanyasweeney.org/work/identifiability.html for Latanya Sweeney's work on de-anonymizing US medical data using date of birth, gender, and zip codes.}. \iffullversion
    Though doubts and concerns about anonymization techniques were voiced well before these works (e.g., \cite{Sweeney1996AnonymizingText}), the
\fi
demonstration of viable attacks was what prompted stakeholders to take the risks more seriously, and catalyzed change in community practices in industry, research, and beyond. (Though these changes were positive, of course, the publishing of the data could not be undone.)\footnote{The tendency to react after a notable harm occurs is certainly not unique to technology policy issues.}

On research methodologies and norms, we learned how to investigate compromised datasets ethically~\cite{Bonneau2012Password}, while taking into account privacy and anonymity considerations and other potential harms, and studied whether and how to use of datasets of illicit origin for research purposes at all~\cite{Thomas2017DatasetOrigin}. 

On modeling, Dwork et al.~\cite{dwork06singleauth,dwork06diffp}\iffullversion, following on privacy-preserving work before them~\cite{dinur03revealing, dwork04datamining},\fi  
~introduced the notion of \emph{differential privacy} (DP), a notion which has also featured impactfully in machine learning research (e.g.,~\cite{YuNBGIKKLMWYZ24,XianLKZ24}). These works discuss DP from a theoretical standpoint, including explanations for modeling decisions, and the DP framework enables precise mathematical reasoning and tradeoff-making about certain kinds of deanonymization risks.

On awareness and adaptation, deanonymization research since the late 1990s gradually impressed upon computer scientists and the broader public that anonymization techniques are unreliable and poorly understood---and later on, that not a single one provided perfect anonymity under researchers' scrutiny, and anonymization might be a pipe dream~\cite{Ohm2009Anonymization, Hern2019AnonymizedData, Biden2024SensitiveData, FTC2024Hashing}. Organizations now appear less likely to optimistically publish datasets that would be damaging if deanonymized. Subsequent works explore whether and how anonymization techniques could be useful despite acknowledged imperfections, if tailored to specific 
\iffullversion
    application contexts and threat models~\cite{Angiuli2016deidentify}.
\fi

\iffullversion
    The community has not been able to reach the elusive ideal of perfect anonymization. 
    Through iterated and transparent research about anonymization's flaws, though, the community has been able to embrace the idea that this goal is infeasible, and to handle datasets accordingly. 
\fi

\section{Novel Challenges}\label{sec:differences}

\iffullversion
    This section highlights significant differences between the domains of security and AI that may necessitate new approaches to transparency in AI.
\fi

\iffullversion
    \subsection{Training Data}\label{sssec:discussion:datatypes}
\fi

In security,
there is usually a clear delineation between (1) a system's design and functionality and (2) (private) input data within the system. In AI, this line is blurred, with the training data for a model being inextricably linked to its performance and properties. Just the untrained model is insufficient for analyzing its properties \cite{Felzmann2020TbyDesign}. Additionally, some key motivations for transparency, such as model bias~\cite{Suresh2021HarmSources} and memorization~\cite{carlini2023extracting}, contend 
explicitly
with training data and its expressions in the trained model. 

Some AI models could reasonably make their training data public. However, if any or all of this data is private or should not be publicly accessible,
disclosing the training set may expose developers and researchers to serious legal and ethical
\iffullversion
    risks.\footnote{This is the case for most useful datasets. See~\cite{Thomas2017DatasetOrigin} for related discussion.} On the other hand, completely obscuring this data may impede important transparency interests such as those discussed in earlier sections.
\fi
To mitigate these concerns, some have proposed disclosing only (partial) model weights or allowing only black-box access, though each of these avenues still carries a significant risk of exposing training data~\cite{Nasr2019Privacy,nasr2023scalableextraction}. Some have proposed anonymized or synthetic training data; these approaches carry their own risks~\cite{Narayanan2008deanon,Cristofaro24}.

Security has not had to face the same issue. 
\iffullversion
    While, e.g., password and biometric research may interface with private data, these may be isolated and analyzed in a more closed-loop setting to avoid unforeseen leakages~\cite{Bonneau2012Password}, and the private data is still mostly conceptually separate from system functionality. 
\fi
We believe this is an important open problem to solve in pursuit of AI transparency, and one in which collaboration between security and AI researchers may be a fruitful avenue to understanding training-data confidentiality concerns and how they interact with transparency~\cite{carlini2023extracting, nasr2023scalableextraction}.

\iffullversion
    \subsection{Disclosure Processes}
\fi

It is important to consider how the disclosure processes may be adapted for the AI context. 
\iffullversion

    In security, the affected parties of a security vulnerability are generally fairly clear in scope. 
    However, as we discuss above, the entanglement of training data with system design makes the relevant stakeholder set larger and potentially unclear. For example, if a vulnerability to memorization attacks is detected in a model~\cite{carlini2023extracting}, should the group making the model be disclosed to, or the owner of that data, or the data subjects, or a combination of the above? This is especially complicated as the interests of these various stakeholders may be in tension.\footnote{Incentives are complicated in disclosures about security too, as system owners typically weigh risk with profit, usability, and other interests beyond just security. We see this, for example, in data breaches~\cite{Cavusoglu2014breach}.}
\fi

While important initial progress has been made on frameworks for AI vulnerability disclosure~\cite{nist2023riskframework, Cattell2024flaw}, the risks and at-risk parties implicated by different vulnerabilities are highly context-dependent. Understanding and careful modelling of what is and is not inherent to AI and machine learning techniques would help piece through these, but it is possible the scope of vulnerabilities and thus responsible disclosure in AI may be an inherently more complex problem.

\iffullversion
    \subsection{Brittleness of Models}
\fi

The well-documented and sometimes inherent tradeoffs between privacy (of training data) and performance is a key feature of AI development~\cite{Carvalho2023TradeOffs}. 
In security, there is almost always a tradeoff between efficiency and privacy, and this
is often the main tradeoff to consider.\footnote{There are contexts in security with more complex tradeoffs, e.g., encrypted search~\cite{Curtmola2006searchable, boneh2015ORE} and differential privacy~\cite{dwork06diffp, Dwork2014DiffPriv}.}
Choosing the right balance of privacy and performance is an inherently non-technical problem~\cite{Schneier2003beyond}, and sometimes,
development toward privacy and performance fundamentally contradict each other~\cite{Gu2022TradeOffs}.
The opportunity for technical work is in the modeling of the
\iffullversion
     tradeoff---one must have an understanding of the tradeoff space to the extent that it is quantifiable and how to push its boundaries for given applications in order to effectively solve the above non-technical question. 
     
\fi
While there are parallels between the two communities, it seems the tradeoff problem in the AI community is certainly different in scale and possibly different in kind (e.g., involving multidimensional optimization ~\cite{Monterio2023MultiOptimization}), posing a greater challenge in finding an appropriate tradeoff---and heightening the possibility of domino effects of accidental repercussions. 

\iffullversion
    \subsection{Optimizing for Metrics}
\fi

    Many AI systems are based on optimizing for metrics serving as heuristics for a real-world objective. Metrics feature across development, from initial training to the addition of post-hoc ``guardrails.''

    A well-established adage from economics (Goodhart's Law) states: ``When a measure becomes a target, it ceases to be a good measure.''~\cite{Goodhart1975Money} In other words, once a metric is known to be used for consequential decisions, it will be gamed--and thus, likely no longer be a good metric.
    In this context, disclosing metrics can make them more of a ``target'' and/or easier to game. These observations would seem to counsel against wide disclosure of metrics (such as objective functions) in order to preserve the utility of the metrics and thus the utility of the system overall.

Yet metrics, like training data, constitute an essential part of system functionality. Analyzing a system given its description with the metrics redacted is likely to result in a significantly incomplete understanding of its functionality. And some key motivations for transparency, such as model bias~\cite{Carlson2017Transparency}, explicitly contend with the metrics used.
As such, the security-by-transparency approach would suggest that transparency of metrics yields important benefits.

The idea that releasing the full details of system functionality can inherently make the system \emph{less useful} for its intended purpose \iffullversion, to our knowledge, \fi does not have a good analogue in security \iffullversion research\fi. 

\section{Perceived Challenges}\label{sec:discussion}

We now briefly discuss arguments against transparency in AI which have parallels to those in security.

\subsection{Trade Secrets and Innovation} AI systems are built on algorithms and training data, the details of which may be important for commercial success. One common concern is that too much disclosure of algorithms and data would reduce incentives for companies to innovate through costly investments in AI (e.g.,~\cite{Hind2020anecdotes,FehrSurvey})---thus harming quality of AI overall as well as competitiveness in a global market.\footnote{For an overview of trade secrecy and innovation, see~
\cite{WIPOTradeSecrets}.} Likewise, industry stakeholders have long used intellectual property and innovation as reasons to argue against transparency of technology in security contexts (e.g., \cite{Chakraborty2009IP}). 

The security research and policy communities have pushed back with counter-arguments ranging from technical to ethical and legal (e.g., 
\cite{Levine2006TradeSecrets, Jones2007RightToKnow}). Many strands of this ongoing ``transparency vs. innovation'' debate in security touch on issues also pertinent to the analogous ongoing debate in AI; and indeed, we see some of the same issues feature in recent commentary on ``transparency vs. innovation'' in AI (e.g., \cite{Ferrandis2022AIIP,Foss-Solbrekk2023TradeSecretLaw, Rowe2022AlgTransparency, Tschider2021BlackBox}).\footnote{We see ``transparency vs. innovation'' as a false dichotomy.}

While the courses of the two debates, and how much they will diverge, remains to be seen, the parallels may be instructive. Notably, significant trends in favor of transparency in the security context have grown despite this kind of pushback, and do not appear to have crippled the industry.

\subsection{Misuse} Another common concern is that releasing AI systems may lead to bad consequences from their deliberate or accidental use for harm, and that the potential or likelihood of such harm may increase with fuller disclosure of how the technology works 
\iffullversion
    (e.g.,~\cite{Brundage2018AIMalUse,Google2019Governance,Gade2024BadLlama}).\footnote{Further, some fear an existential risk to wide dissemination of AI 
    tools, with this fear sometimes cited as a reason to be wary of transparency~\cite{Mak2019dangerous, Thaler2024dangerous, Vincent2019dangerous}.
    Such fears do not have an analogue in security.}
\fi

These narratives have limited parallels to long-standing discussions in the security community around privacy tools (such as encrypted messaging~\cite{Jarvis2020Wars} and Tor~\cite{Dingledine2004Tor}) being used by bad actors and for illegal activity.
The security community has overwhelmingly advocated, over the course of these debates, that it is critical to research and develop these kinds of general-purpose or so-called ``dual-use'' security and privacy tools~\cite{Silic2013dualuse, OTA1993dualuse}, both because the exact same technical functionality that can be misused also provides critical protection for dissidents, journalists, victims, and others who use it for good~\cite{Amnesty2016Encryption, HRC2022RightToPrivacy}, and because only by studying the topic can we better understand the harms and how to mitigate them while promoting the above benefits. 

Of course, this determination is context-dependent. 
\iffullversion
    While the AI community's approach may take time to crystallize---as did the security community's, and as does any complex community consensus building---aspects of the thought process around dual-use privacy tools may have useful parallels to concerns about misuse in AI.
\fi

\smallskip


\iffullversion
    \subsection{Pace of Development \\ and Competitiveness} 
    Many have commented on the unusually fast pace of recent AI developments, often voicing concerns that it is moving too fast (e.g.,~\cite{Bengio2024ManagingRiskSpeed,Tang2020AISpeed}). 
    This pace has naturally impeded the establishment of standards and best practices ~\cite{Bostrom2017OpenDev} and limited the ability of researchers and developers to reach fuller understandings of of new techniques before publication or deployment, as well as the
    drafting and implementation of timely regulation~\cite{Ruschemeier2023AIReg}.
    
    In our view, while the security community has not experienced a comparable pace of development, it has still slowed down compared to an earlier phase, 
    based on a collective experience of flaws being discovered in secure systems and cryptographic designs  that were developed too quickly and scrutinized insufficiently before deployment (some of which caused major security problems with serious impact, e.g.,~\cite{Long2009SoftwareGlitch}). Over time, the community moved towards investing much more research and development in precisely defining context-specific security requirements, extensive testing of new systems, models, and assumptions before relying on them, and to widely regard as unreliable systems lacking this kind of development and testing over time. The time-consuming step of planning ahead in detail for the security failures that will inevitably occur even if all of these precautions are taken also became more common over time. 
    
    Hence, expectations around pace of development and what makes a product competitive shifted, in line with a widespread belief (and lived experience) that more deliberation leads to better products and rushed development leads to real risks.
    The security community's experience, for what it is worth, illustrates a gradual, costly collective agreement that taking time for transparency does not curb innovation, but rather promotes it, and the delays are worth it to prevent disasters.
\fi

\iffullversion
    \subsection{Consequential Outcomes \\ and a Right to Know} As AI systems are used in ways that have  increasingly consequential outcomes,\footnote{E.g., in healthcare \cite{Khan2023HealthcareAI}, child custody \cite{Brooks2022AIFamilyMatters}, or bail \iffullversion{setting} \fi \cite{Fine2024BailTrustAI}.} some calls for transparency center the idea of a ``right to know'' about decisions that impact individuals or groups (e.g., \cite{Fehr2024TrustworthyAI}). In the European Union, the AI Act has recently written certain aspects of a ``right to know'' into law \cite{EU2024AIAct}. The reasoning underlying these calls generally focuses broadly on the use of (opaque) algorithms in consequential decision-making procedures, and as such is not unique to AI, although modern ML may exacerbate these concerns due to its complexity and the current trend of ``black-box'' use \cite{Gryz2021BlackBoxRights}. Quite naturally, then, similar discussions have arisen in the context of security-critical systems that impact consequential outcomes (e.g., in election security and government technology~\cite{Girgvliani2023procurement,Jones2007RightToKnow}). 
\fi

\iffullversion
        \section{Other Related Work}\label{sec:related}
    
    This work is concerned with parallels between transparency in security and AI. Of course, there are many different fields which overlap with these ideas. 
    
    \subsection{Comparison To Other Communities}\label{sec:communities}
    
    \paragraph{Biosecurity} Sam Marks, in referencing Jeff Kaufman's discussion~\cite{Kaufman2023ComCulture} on differences in norms between the biosecurity and cybersecurity communities, questions if AI would be better suited to a comparison to biosecurity~\cite{Marks2023OSAI}. Primarily, this is due to similar difficulties in fixing vulnerabilities. This comparison would frame AI as more of a governmental responsibility than currently understood. 
    
    \paragraph{Physical Sciences} There has been a recent uptick in discussion about data documentation in the Physics community~\cite{Chen2024PhysicsRetractions}. This has raised numerous concerns about the standards of trust for auditors and researchers alike. Many are calling for strengthening ethical standards, something they say can never stop being developed~\cite{Houle2024PhysicsTrust}. 

    Issues of ethics and public trust in nuclear physics also have some limited parallels to discussions around ethics and transparency in AI, and have been compared to ethical issues in cryptography~\cite{rogaway2015ethics}.
    
    \paragraph{Biomedical Sciences} The pharmaceutical industry is a community that requires collaboration across many stakeholders and relies strongly on public trust. They have a history of strong risk frameworks in regards to societal impact~\cite{Pire-Smerkanich2016PharmaFrameworks} and are heavily regulated by governmental bodies. Notably, there is a documented increase in public trust when transparency is valued in the industry~\cite{Singh2023PharmaTrust}. 
    
    \paragraph{Civil Engineering} The civil engineering community has widespread responsibility in ensuring public safety. Therefore, they have well-established standards of measurability (outlining specs, etc.) as well as a well developed auditability processes~\cite{ASCE2024Standards}. Here, we see correlation between public trust and transparency in the community~\cite{Crumpacker2008PublicTrust}.
    
    \subsection{Related concepts}
    
    \paragraph{Securing AI} While we are discussing relationships between security and AI, our focus is not on \emph{securing AI}---that is, how to build AI systems that are secure. However, many efforts in modelling and implementing transparent AI contend explicitly with preventing harm and security risks. This includes privacy for training data through, e.g., differential privacy~\cite{Dwork2014DiffPriv} and unlearning~\cite{Garg2024Unlearning}; preventing misclassification and misuse through, e.g., the study of adversarial robustness~\cite{Carlini2019Adversarial,Awasthi2023Adversarial}; and labelling of AI generations through watermarking~\cite{Christ2024PrECC}.
    
    \paragraph{Other Notions of Transparency in AI}
    Prior work has offered taxonomies of transparent AI from 
    various perspectives, such as frameworks for AI risk~\cite{NIST2023Framework}, trustworthy AI~\cite{Newman2023TrustworthyAI}, ethics guidelines~\cite{EU2019AIEthics}, and compilations of previous works~\cite{Graziani2022Taxonomy}.
    
    \paragraph{Red-Teaming and Jail Breaking} 
    Red-teaming and jailbreaking AI 
    to test the limits of models
    has become more common, and important recent progress
    has been made
    (e.g,~\cite{Wei2023Jailbreak, Carlini2024Stealing, Rando2022RedTeaming, Feffer2024RedTeaming, Longpre2024RedTeaming}).
    
    \section{Conclusion}
    Transparency is a key concept, and has been a subject of community controversy, in both security and AI.  
    We explain how the current discourse on AI transparency exhibits informative parallels with past and ongoing discourse on transparency in security, and discuss three key parallels---modelling (\S\ref{ssec:mainarg:modeling}), community input (\S\ref{ssec:mainarg:disclosure}), and public trust (\S\ref{ssec:mainarg:interpretability})---illustrated with a case study (\S\ref{ssec:casestudy:deanon}). We systematize common themes in arguments against transparency in both domains (\S\ref{sec:discussion}), and highlight novel challenges presented by transparency in the context of AI (\S\ref{sec:differences}).
\fi

\ifanon\else
    \iffullversion
        \section*{Acknowledgments}
    \else  
        \section{Acknowledgments}
    \fi
    We are grateful to
    Kyunghyun Cho,
    Ana-Maria Cretu,
    Andr\'{e}s F\'{a}brega,
    Betty Li Hou, 
    Robert Mahari,
    Falaah Arif Khan,
    Adam Sealfon,
    Vitaly Shmatikov,
    and our anonymous reviewers at AAAI 2025
    for helpful discussions and feedback on drafts.
\fi

\printbibliography

\end{document}